\documentclass[twocolumn,showpacs,preprintnumbers,amsmath,amssymb]{revtex4}
\usepackage{graphicx}
\usepackage{dcolumn}
\usepackage{bm}

\begin{document}

\preprint{PRL 99 (2007) 150404}

\title{Nonlinear phase shift from photon-photon scattering in vacuum}

\author{Albert Ferrando,$^{1}$ Humberto Michinel,$^{2}$
Marcos Seco,$^{3}$ and Daniele Tommasini$^{2}$ }

\affiliation{$^{1}$Interdisciplinary Modeling Group, InterTech. Departament d'\`{O}ptica,
Universitat de Val\`{e}ncia. Dr. Moliner, 50. E-46100 Burjassot (Val\`{e}ncia),
Spain.}

\affiliation{$^{2}$Departamento de F\'\i sica Aplicada. Universidade de Vigo.
As Lagoas E-32004 Ourense, Spain.}

\affiliation{$^{3}$Departamento de F\'{\i}sica de Part\'{\i}culas. Universidade de Santiago de Compostela.
15706 Santiago de Compostela, Spain.}


\begin{abstract}
We show that QED nonlinear effects imply a phase correction to the linear evolution
of electromagnetic waves in vacuum. We provide explicit solutions of the modified Maxwell's
equations for the propagation of a superposition of two plane waves, and calculate
analytically and numerically the corresponding phase shift. This provides
a new framework for the search of all-optical signatures of photon-photon
scattering in vacuum. In particular, we propose an experiment for measuring
the phase shift in projected high-power laser facilities.
\end{abstract}

\pacs{12.20.Ds, 42.50.Xa, 12.20.Fv}

\maketitle

{\em Introduction.-} Quantum Electrodynamics (QED) predicts that
at photon energies well below the electron rest energy, photon-photon
collisions can still be produced through the interchange of virtual
electron-positron pairs\cite{Halpern,Heisenberg}. This nonlinear interaction
modifies Maxwell's equations for the average values of the electromagnetic
quantum fields\cite{mckenna} and affects the properties of the QED vacuum\cite{klein}.
During many years, the search of these effects has been restricted to
projected particle physics experiments with accelerators. However, photon-photon
scattering processes in vacuum will become testable at energy densities
achievable with ultrahigh power lasers in the near future\cite{mourou06}.

The realization of these ultra intense photon sources began with the
discovery of chirped pulse amplification (CPA)\cite{strickland85} in the
late 80's and optical parametric chirped pulse amplification (OPCPA)
\cite{dubietis92} in the 90's. These techniques opened the door to a
field of research in the boundary between optics and experimental
high-energy physics, where lots of novelties are expected to
come in the next years. In fact, several recent works propose different configurations
that can be used to test the nonlinear optical response of the vacuum, e.g.
using harmonic generation in an inhomogeneous magnetic field\cite{ding92}, QED
four-wave mixing\cite{4wm}, resonant interactions in microwave cavities\cite{brodin},
or QED vacuum birefringence\cite{Alexandrov} which can be probed by x-ray pulses\cite{xray},
among others\cite{others}.

In the present Letter, we show that QED nonlinear effects in vacuum imply a
phase correction to the linear evolution of crossing electromagnetic waves. We provide
explicit numerical and analytical approximate solutions for the propagation of
a superposition of two plane waves. This result allows us to calculate the corresponding
phase shift, providing a new framework for the quest of signatures of photon-photon
scattering. In particular, we suggest an experiment for measuring the effect of nonlinear
vacuum in projected high-power laser facilities like the European Extreme Light Infrastructure
(ELI) project\cite{ELI} for near IR radiation.

{\em Model and equations-.} Let us begin by writing the formulae that have to be
used instead of the classical linear Maxwell equations, including the terms which come
from QED effects in vacuum. The corresponding Lagrangian density in terms of
the electric and magnetic fields {\bf E } and {\bf B}, was derived in the 30's
by Euler and Heisenberg\cite{Heisenberg}:

\begin{equation}
{\cal L}={\cal L}_0+\xi{\cal L}_Q={\cal L}_0
+\xi\left[{\cal L}_0^2+ \frac{7\epsilon_0^2 c^2}{4}({\bf E}\cdot {\bf B})^2\right],
\label{L}
\end{equation}
being
\begin{equation}
{\cal L}_0=\frac{\epsilon_0}{2}\left({\bf E}^2-{c^2\bf B}^2\right)
\label{L0}
\end{equation}
the linear Lagrangian density and $\epsilon_0$ and $c$ the dielectric
constant and the speed of light in vacuum, respectively. As it can be
appreciated in Eq. (\ref{L}), QED corrections are introduced by the parameter
\begin{equation}
\xi=\frac{8 \alpha^2 \hbar^3}{45 m_e^4 c^5}\simeq 6.7\times
10^{-30}\frac{m^3}{J}.
\label{constant_xi}
\end{equation}
This quantity has dimensions of the inverse of an energy density. This means that
significant changes with respect to linear propagation can be expected for values around
$\vert\xi{\cal L}_0\vert\sim 1$, corresponding to beam fluxes with electromagnetic energy densities given by
the time-time component of the energy-momentum tensor
\begin{equation}
T_{00}=\frac{\partial{\cal L}}{\partial (\partial_tA)}\partial_t A - {\cal L}
\gtrsim 2/\xi\simeq 3\times 10^{29} J/m^3.
\label{T00}
\end{equation}
While such intensities may have an astrophysical or cosmological importance,
they are not achievable in the laboratory. The best high-power
lasers that are being projected for the next few decades will be
several orders of magnitude weaker, eventually reaching energy
density of the order $\rho\sim 10^{23} J/m^3$\cite{mourou06}. Therefore, we will
study here the ``perturbative" regime, in which the non-linear
correction is very small, $\vert\xi{\cal L}_0\vert \ll 1$. As we
shall see, even in this case measurable effects can be
accumulated in the phase of beams of wavelength $\lambda$ traveling over
a distance of the order $\lambda \vert\xi{\cal L}_0\vert^{-1}$. Thus, current sensitive
techniques could be used to detect traces of QED vacuum nonlinearities.

Once the electromagnetic fields are expressed in terms of the four-component gauge
field $A^\mu=(A^0,{\bf A})$ as ${\bf B}=\nabla \wedge {\bf A}$ and
${\bf E}=-c \nabla A^0-\frac{\partial {\bf A}}{\partial t}$, the equations
of motion are given by the Variational Principle:

\begin{equation}
\frac{\delta \Gamma}{\delta A^\mu}=0,
\label{var_princ_gen}
\end{equation}
where $\Gamma\equiv \int {\cal L} d^4x$ is the QED effective
action. Instead of studying the resulting equations for the fields
${\bf E}$ and ${\bf B}$, that can be found in the literature\cite{mckenna,variational},
for the present purposes it is more convenient to consider the equations for the gauge field components $A^\mu$.
In general, these four equations cannot be disentangled. However, after some straightforward algebra
it can be seen that they admit solutions in the form of linearly polarized
waves, e.g. in the $x$ direction, with $A^0=0$ and ${\bf A}=(A,0,0)$, provided
that: $i)$ the field $A$ does not depend on the variable $x$ (a {\em transversality}
condition) and $ii)$ $A(t,y,z)$ satisfies the single equation:

\begin{equation}
\partial_\mu\partial^\mu A + \xi\epsilon_0\left(\partial_\mu\partial^\mu A\partial_\nu A\partial^\nu A+
2\partial_\mu A\partial_\nu^\mu A\partial^\nu A\right)=0,
\label{covariant}
\end{equation}
where we have used the convention $g_{\mu\nu}={\rm diag}(1,-1,-1,-1)$ for the metric tensor.
Hereafter, we will restrict our discussion to this case of linearly-polarized solution.
The orthogonality relation ${\bf E}\cdot {\bf B}=0$ are then automatically satisfied,
and the effective Lagrangian Eq. (\ref{L}) reduces to
${\cal L}={\cal L}_0\left(1+ \xi{\cal L}_0\right)$. Note that the plane-wave
solutions of the linear Maxwell equations,
such as ${\cal A}\cos\left({\bf k}\cdot{\bf r}- \omega t\right)$,
where ${\cal A}$ is a constant, ${\bf k}=(0,k_y,k_z)$ and $\omega=c \vert {\bf k}\vert$,
are still solutions of Eq. (\ref{covariant}). However, we expect
that the non-linear terms proportional to $\xi$, due to the QED correction,
will spoil the superposition principle.

{\em Variational solutions-.} Let us consider two plane waves,
for simplicity having the same phase at the space-time origin,
having wave vectors $(0,q,k)$ and $(0,-q,k)$ respectively, and angular
frequency $\omega=\sqrt{k^2+q^2}$. Any of them, when taken alone, would be a
solution of both the linear and non-linear equations. However, the
superposition

\begin{eqnarray*}
A(t,y,z)& = &\frac{\cal A}{2}\left[\cos(k z- \omega t+q y)+\cos(k z-
\omega t - q y)\right]\\ \nonumber
& = & {\cal A}\cos(q y)\cos(k z - \omega t) ,
\end{eqnarray*}
where ${\cal A}$ is a constant, would only solve the linear
equations of motion. In our perturbative regime, we can expect
that the small non-linear correction will progressively modify the
form of $A(t,y,z)$ as the wave proceeds along the $z$ direction.
We will therefore make the ansatz
\begin{equation}
A = {\cal A}\cos(q y)\left[\right.\alpha(z)\cos(k z - \omega t) +
\beta(z)\sin(k z - \omega t)\left.\right],
\label{Axpol_nonlin}
\end{equation}
allowing for the generation of the other
linearly-independent function $\sin(k z - \omega t)$
(we will take $\alpha(0)=1$ and $\beta(0)=0$).
On the other hand, we neglect the possible generation of a reflected wave
depending on $k z + \omega t$, which can be expected to be a
smaller correction in this perturbative regime.
As we will discuss below, in our perturbative regime the effects of
the possible $y$-dependence of $\alpha$ and $\beta$ are negligible
for the traveling distances in the $z$ direction that we will consider.

According to the Variational Method, we require that the functions
$\alpha(z)$ and $\beta(z)$ correspond to a local minimum of the
effective action $\Gamma$, after averaging out $y$ and $t$ as follows:

\begin{equation}
\Gamma =\int_{-\infty}^{\infty}d z \left(\frac{q \omega
}{4\pi^2}\int_{-\pi/q}^{\pi/q}dy \int_{0}^{2\pi/\omega}dt {\cal
L}\right),
\label{Gamma_constraint}
\end{equation}
whose minimum corresponds to the equations $\delta \Gamma/\delta
\alpha=0$ y $\delta\Gamma/\delta \beta=0$. After a straightforward
computation, these two equations can be written as

\begin{eqnarray}
\label{var1}
&& \frac{\alpha ''}{2} + k \beta' + a q^4 \alpha (\beta ^2 + \alpha ^2)
-\frac{a}{16}\left[2 \left(9 k^2-4 q^2\right) \beta ^2\right. \\ \nonumber
&&  \left.\;\;\;\;\; +54 k \beta \alpha '+27 \alpha '^2 + 6 k^2
\alpha ^2 - 18 k \alpha\beta' + 9
\beta '^2 \right] \alpha ''\\ \nonumber
&& +\frac{a}{8}\left[9 k \alpha\alpha' -9 \alpha' \beta'-9 k \beta
\beta' + 2 \left(3 k^2-2 q^2\right) \alpha \beta \right]
\beta ''\\ \nonumber
&&  +\frac{a}{8}\left[-9 k \beta '^3  - 3 k^2 \alpha  \alpha '^2 +
\left(15 k^2-8
q^2\right)\alpha  \beta '^2 \right] \\ \nonumber
&&-\frac{a}{8}\left[9 k \alpha '^2- 16 k q^2\alpha ^2
-16 k q^2 \beta ^2+ 2 \left(9 k^2-4 q^2\right)\alpha' \beta  \right]\beta ' \\ \nonumber
&&  =0,
\end{eqnarray}
and

\begin{eqnarray}
\label{var2}
&& \frac{\beta ''}{2} - k \alpha' + a q^4 \beta (\alpha ^2 + \beta
^2)
-\frac{a}{16}\left[2 \left(9 k^2-4 q^2\right) \alpha ^2\right.\\ \nonumber
&& \left. \;\;\;\;\;-54 k \alpha \beta '+27 \beta '^2 + 6 k^2
\beta ^2 + 18 k \beta\alpha' + 9
\alpha '^2 \right] \beta ''\\ \nonumber
&& +\frac{a}{8}\left[-9 k \beta\beta' -9 \beta' \alpha' + 9 k
\alpha \alpha' + 2 \left(3 k^2-2 q^2\right) \beta \alpha \right]
\alpha ''\\ \nonumber
&& +\frac{a}{8}\left[9 k \alpha '^3  - 3 k^2 \beta  \beta '^2 +
\left(15 k^2-8
q^2\right)\beta  \alpha '^2 \right]\\  \nonumber
&&-\frac{a}{8} \left[-9 k \beta '^2 + 16 k q^2\beta ^2
+16 k q^2 \alpha ^2+ 2 \left(9 k^2-4 q^2\right)\beta' \alpha  \right]\alpha ' \\ \nonumber
&& =0,
\end{eqnarray}
where $a=\xi \epsilon_0{\cal A}^2 / 2$.
It is now convenient to express the parameter $a$ in terms of the energy density,
\begin{eqnarray}
\label{denergy}
\rho &=& T_{00}\\ \nonumber
&=& \frac{\epsilon_0}{2}({\bf E}^2+c^2{\bf B}^2)+\frac{\xi}{4}
\epsilon_0^2 ({\bf E}^2-c^2{\bf B}^2)\left(3 {\bf E}^2+c^2{\bf
B}^2\right).
\end{eqnarray}
In our perturbative regime,
this gives $a\simeq\frac{2 c^2}{\omega^2}\xi\rho$
with a very good approximation.

{\em Numerical simulations and approximate analytical solution-.} The result
of the numerical integration of Eqs. (\ref{var1}) and (\ref{var2})  is
shown in Fig. 1 for a choice of parameters that may be accessible
at future facilities\cite{mourou06}, namely: $\rho=4\times 10^{23} J/m^3$, $\lambda=5\times 10^{-7} m$,
$k=q=\frac{2\pi}{\sqrt{2}\lambda}=0.89\times 10^7 m^{-1}$, giving $a=3.4\times 10^{-20} m^2$.
The two enveloping functions show a
sinusoidal behavior, with an oscillation length $\sim 0.26 m$.
In particular, we find that a 100\% change in phase is accumulated after
the distance $\Delta z\simeq 13 cm$.

\begin{figure}[htb]
{\centering \resizebox*{1\columnwidth}{!}{\includegraphics{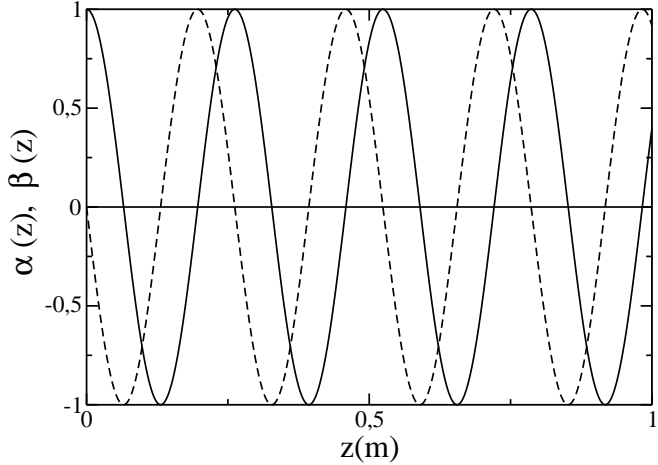}} \par}
\caption{Numerical solution of Eqs. (\ref{var1}) and (\ref{var2})
for $k=q=0.89\times 10^7 m^{-1}$ and $a=3.4 \times 10^{-20} m^2$.
The continuous curve represents the function
$\alpha(z)$, the dashed curve the function $\beta(z)$.}
\label{fig1}
\end{figure}

This numerical result also suggests the viability of
an analytical approximation. In fact, the numerical oscillations show that each order in derivation
of $\alpha$ and $\beta$ corresponds to a suppression factor
$\sim 2\pi/0.26 m^{-1}$, 6 orders of magnitude smaller than $k$ and $q$.
We can then neglect the second derivatives in Eqs. (\ref{var1}), (\ref{var2}).
Moreover, in our perturbative regime $a$ can also be considered an expansion parameter,
therefore we can also neglect the first derivatives when they appear multiplied by $a$.
As a result, Eqs. (\ref{var1}), (\ref{var2}) can be approximated as follows:

\begin{equation}
k \beta' + a q^4 \alpha (\alpha ^2+\beta ^2)=0,
\end{equation}

\begin{equation}
- k \alpha' + a q^4 \beta (\alpha ^2 + \beta ^2)=0,
\end{equation}
whose analytical solution is

\begin{equation}
\alpha(z)=\cos\left(\chi z\right), \;\;\; {\rm and} \;\;\;
\beta(z)=-\sin\left(\chi z\right),
\end{equation}
where

\begin{equation}
\chi\equiv a q^4/k=\xi \epsilon_0{\cal A}^2 q^4/2k\simeq 2 \frac{c^2 q^4}{k\omega^2} \xi \rho.
\label{nlphase}
\end{equation}

This result fairly coincides with our numerical solution, giving $\chi=24 m^{-1}$
and $2\pi/\chi=0.26 m$ for the same choice of the parameters as in Fig. 1.
This is an {\it a posteriori} justification for the analytical approximation.

By substituting in Eq. (\ref{Axpol_nonlin}) and after some trigonometry, we find
that our approximated {\it analytical} solution of the non-linear
equations for the electromagnetic field in the vacuum can be
written as

\begin{equation}
A(t,y,z)= {\cal A}\cos(q y)\cos(k z + \chi z- \omega t) .
\label{Axpol_nonlin2}
\end{equation}

Therefore, the effect of the non-linearity is to change the phase
of the wave, with respect to the linear solution, by a term that
increases linearly with the distance. This can also be interpreted
as a change in the $z$ component of the wave vector, which becomes
$k_z=k+\chi$, so that the dispersion relation is modified to
$\omega=c\sqrt{(k_z-\chi)^2+q^2}$.

To conclude this section, let us discuss now the validity of the approximations that we have made.
The Variational Method that we have used is expected to provide very good results whenever the class of test functions,
given by Eq. (\ref{Axpol_nonlin}) in our case, is a reasonably good choice.
Since the latter was motivated by perturbative considerations, the whole approximation
will be justified whenever the correction $\chi\ll k$, i.e. whenever $2 \frac{c^2 q^4}{k^2\omega^2} \xi \rho\ll 1$.
This is guaranteed by the dispersion relation and the fact that $\xi \rho\ll 1$ in our regime.
It is easy to see that in this case the correction to the energy density dependence on $k$ and $q$ is also very small.

The other approximation that we have made was neglecting the $y$-dependence of the enveloping functions $\alpha$ and $\beta$.
For an {\it a posteriori} test of this hypothesis, we have corrected our solution
Eq. (\ref{Axpol_nonlin2}) allowing for the further envelop functions $\gamma(y,z)$
and $\sigma(y,z)$. We have then used as a new ansatz for the Variational Method the potential
$A\equiv {\cal A}\cos(q y) (\gamma \cos(k_z z - \omega t)+\sigma \sin(k_z z - \omega t))$,
where $k_z=k+\chi=k+ a q^4/k$ from Eq. (\ref{nlphase}).

We have performed a first average over the fast variation in $y$ in the action
due to the trigonometric dependence on the product $q y$, and we have minimized
the action. After keeping only the first order in the expansion parameter $a$,
and neglecting the terms involving the second derivatives (which imply a slower
variation as discussed above), we have got the equations
$\partial_z\sigma +  \chi\gamma\left( \gamma^2+\sigma^2-1\right)=0$,
and $\partial_z\gamma - \chi\sigma\left( \gamma^2+\sigma^2-1\right)=0$.
These equations, which do not involve $\partial_y\gamma$ and $\partial_y\sigma$
at the first order, are solved by $\gamma(y,z)=1$ and $\sigma(y,z)=0$, therefore
we conclude that at this order our previous solution, Eq. (\ref{Axpol_nonlin2}), is not modified.
This justifies the approximation of neglecting the possible $y$-dependence in the first instance.

{\em Comparison with the optical Kerr effect-.} Since it is theoretically known that the
vacuum shows birefringence as in the DC Kerr effect\cite{klein}, it is interesting to compare
our result with the optical (AC) Kerr effect that also arises in matter. In fact, let us
consider a Kerr medium, characterized by an effective nonlinear refractive index of the form:

\begin{equation}
n = n_0 + n_2I,
\label{Kerr}
\end{equation}
where $n_0$ is the linear refractive index, $n_2$ the Kerr coefficient
and $I$ the irradiance of the beam. In such a medium,
for propagation through a distance $d z$, the
wavefront phase will be modified by an amount:

\begin{equation}
d\Phi=\frac{2\pi}{\lambda}n_2Idz.
\label{phase_shift}
\end{equation}

Since $\rho=I/c$, Eqs. (\ref{phase_shift}) and (\ref{nlphase}) show that our configuration
of two crossing waves in vacuum undergoes the same phase shift as a single plane wave in a
Kerr medium having $n_0=1$ and a Kerr coefficient $n_2=2 c \xi q^4/(\omega^2 k^2)$. Choosing
e.g. $q=k=\omega/(\sqrt{2} c)$, this Kerr coefficient would be $n_2=\xi/(\sqrt{2} c)\approx 10^{-38}m^2W^{-1}$.
However, in spite of this analogy, it is important to note that such an $n_2$ cannot be
interpreted as a Kerr coefficient for the vacuum. In fact, strictly speaking the vacuum
{\it does not} show the usual AC Kerr effect. In a Kerr medium, the phase shift (\ref{phase_shift})
is found even for a single plane wave propagating along the $z$-direction. In vacuum, such a
single plane wave, corresponding to $q=0$ in Eq. (\ref{nlphase}), would propagate with $\chi=0$,
i.e. without any phase change, exactly as in the linear case. Ultimately, this is due the fact that
the cross section for photon-photon scattering vanishes for parallel momenta.

{\em Proposal of an experiment-.} Our previous discussion suggests that we
might test the non-linear properties of the QED vacuum by using sensitive
experiments in which the key point is the ability of measuring small phase
changes in a laser beam. In our configuration, a laser pulse is divided in three
beams of the same intensity and two of the resulting beams are focused independently.
The trajectories of both rays cross at the focus point with an angle $\theta$. As a result, the central
part of each distribution has acquired a phase shift $\Delta\Phi$.
In a experiment corresponding to the parameters of the ELI project in its first
step we have pulses of wavelength $\lambda=800nm$, intensity $I=10^{29}Wm^{-2}$ and
duration $\tau=10fs$ which are focused in a spot of diameter $d\approx 10\mu m$.
Either from Eq. (\ref{nlphase}), or equivalently from Eq. (\ref{phase_shift}), this
results in a phase shift $\Delta\Phi\approx 10^{-7} rad$, which can be resolved comparing
with the third beam which was not exposed to the effects of QED vacuum. Current
techniques like spectrally resolved two-beam coupling, which can be applied for ultrashort
pulses\cite{kang97}, can be used to this purpose.
Comparing to other alternatives like x-ray probing of QED birefringence, our
system does not need an extra free electron laser and the power requirements of
the system are only one order of magnitude higher. Moreover, the measurement of
the ellipticity and the polarization rotation angle in birefringence experiments
is not yet possible with current technology. Other techniques like four-wave
mixing processes\cite{4wm} require the crossing of at least three beams, with
the corresponding alignment problems and the rest of the requirements are similar
to our proposal.

{\em Conclusions.-} In conclusion, we have calculated the phase shift arising from propagation of ultra-intense
radiation in vacuum, and shown that it could be measured in the first step of ELI facility under construction.
We consider that the present work could serve as a starting point for the quest of other nonlinear optical phenomena
that may arise in ultra-high power laser beams propagating in vacuum.

{\em Acknowledgments.-} We thank Miguel \'Angel Garc\'{\i}a-March, Pedro Fern\'andez
de C\'ordoba, Davide Tommasini and Mario Zacar\'es
for useful discussions and help. One of the authors (D. T.)
would also like to thank the whole InterTech group at the Universidad Polit\'ecnica de
Valencia for its warm hospitality. This work was partially supported by contracts PGIDIT04TIC383001PR
from Xunta de Galicia, FIS2004-02466, FIS2005-01189 and TIN2006-12890 from the Government of Spain.

\end{document}